\documentclass[11pt,letterpaper,superscriptaddress]{JHEP3}
\usepackage{amssymb,amsmath,amsfonts,amstext,graphics}
\usepackage{cite}
\usepackage{graphicx}
\usepackage{epstopdf}
\input epsf.tex
\newcommand{\beq}{\begin{eqnarray}}
\newcommand{\eeq}{\end{eqnarray}}

\def\ltap{\ \raise.3ex\hbox{$<$\kern-.75em\lower1ex\hbox{$\sim$}}\ }
\def\gtap{\ \raise.3ex\hbox{$>$\kern-.75em\lower1ex\hbox{$\sim$}}\ }
\newcommand{\gsim}{\lower.7ex\hbox{$\;\stackrel{\textstyle>}{\sim}\;$}}
\newcommand{\lsim}{\lower.7ex\hbox{$\;\stackrel{\textstyle<}{\sim}\;$}}
\newcommand{\TRH}{T_{\text{RH}}}
\def\OO{{\cal O}}

\def\tr{{\rm\ Tr}}

\def\be{\begin{equation}}
\def\ee{\end{equation}}
\def\bea{\begin{eqnarray}}
\def\eea{\end{eqnarray}}

\newcommand{\vev}[1]{ \left\langle {#1} \right\rangle }

\newcommand{\kev}{{\rm keV}}

\newcommand{\tev}{{\rm TeV}}

\newcommand{\TeV}{\,\mathrm{TeV}}

\newcommand{\oraf}{O'Raifeartaigh\ }
\def\unit{\relax{\rm 1\kern-.26em I}}
\newcommand{\half}{{\frac{1}{2}}}

\newcommand{\susy}{{\text{susy}}}
\newcommand{\ssb}{{\text{ssb}}}

\title{Reheating Metastable O'Raifeartaigh Models}
\author{Nathaniel J. Craig$^{a,b}$, Patrick J. Fox$^c$ and Jay G. Wacker$^{a,b}$\\
$^a$ 
Theory Group,\\ 
Stanford Linear Accelerator Center,\\
Menlo Park, CA 94025\\
$^b$ Institute for Theoretical Physics,\\ 
Stanford University,\\ 
Stanford, CA 94306\\
$^c$ Theoretical Physics Group,\\
Lawrence Berkeley National Laboratory,\\
Berkeley, CA  94720}
\abstract{
In theories with multiple vacua, reheating to a temperature greater than the height of a barrier can stimulate transitions from a desirable metastable vacuum to a lower energy state.  We discuss the constraints this places on various theories and demonstrate that in a class of supersymmetric models this transition does not occur even for arbitrarily high reheating temperature.
}

\begin{document}

\section{Introduction}

Spontaneously broken supersymmetry (susy) is an attractive solution to the hierarchy problem \cite{Dimopoulos:1981zb}.  The breaking of supersymmetry typically occurs in a sector sequestered from the fields of the MSSM.
In the sequestered sector the minimum of the effective potential cannot satisfy the F-flatness condition for all of the fields. It is not necessary for this minimum to be the global minimum of the potential \cite{Dimopoulos:1997ww,Luty:1997ny,Dimopoulos:1997je} provided that the tunneling rate to the true minimum is slow enough that the Universe has a lifetime longer than its observed age.  
Such metastability appears to be generic when embedding the MSSM into a  larger theory such as string theory.
It is currently believed that the true vacuum of string theory is supersymmetric and therefore phenomenologically unsuitable, but there exists a huge landscape of metastable vacua that are cosmologically long-lived where the SM is realized.

Models of dynamical supersymmetry breaking whose global minimum is non-supersymmetric must satisfy several  constraints  \cite{Witten:1982df,Affleck:1983mk,Nelson:1993nf}.   These constraints are not necessary for theories that have global supersymmetric vacua and a long lived metastable vacuum.  For example, it has recently been observed that massive supersymmetric QCD possesses susy breaking local minima  whose lifetimes can be longer than the present age of the universe \cite{Intriligator:2006dd}. 
Similarly, in \oraf models that have their small mass scales generated through dynamics, or``retrofitted'' \cite{Dine:2006gm},  supersymmetric minima are brought in from infinity and the susy breaking minimum becomes metastable. The simplicity of these theories demonstrates the genericness of metastable minima and suggests that some variant of these theories may provide a phenomenologically viable supersymmetry breaking mechanism.

In the early Universe, after inflation, the visible sector was reheated by inflaton decays to a very high temperature.  The susy breaking hidden sector  can also be reheated to a comparable temperature, for instance if the inflaton decays democratically to all fields or if the inflaton originates in the hidden sector.  Even if the inflaton does not directly reheat the DSB sector, the DSB sector  will be reheated if it is sufficiently well coupled to the visible sector (e.g., in low-scale gauge mediation). At high temperatures the local susy breaking minimum is typically in a different place from the low temperature minima and as the Universe cools, the system may evolve towards either the susy breaking or susy preserving vacuum.   In this note we address this thermal evolution of the effective potential in models with metastable dynamical supersymmetry breaking.

The organization of the paper is as follows.  In Sec.~\ref{sec:reheat} other cosmological constraints on supersymmetry breaking are discussed.  While these constraints can be severe for high reheating temperatures, we describe scenarios where they are not problematic.  In Sec.~\ref{Sec: ISS} we describe the thermal history of Intriligator, Seiberg and Shih (ISS) models \cite{Intriligator:2006dd} of supersymmetry breaking, see also \cite{Schmaltz:2006qs,Forste:2006zc,Banks:2006ma,Braun:2006da,Braun:2006em,Ooguri:2006pj,Garcia-Etxebarria:2006rw,Franco:2006es} for related works.    This theory is a prototypical example of a model that cools into a metastable minimum rather than a supersymmetry preserving one; we elucidate the essential features necessary for this to occur.  In Sec.~\ref{Sec: ORaif} we illustrate how the requirement of cooling into the metastable vacuum may impose additional constraints upon certain classes of hidden sector theories.   In Sec.~\ref{Sec: Early Universe} the effects of moduli trapping and their role in thermalizing the hidden sector are discussed. Finally, in Sec.~\ref{Sec: Conclusion} we discuss the implications for susy breaking and populating a landscape of vacua.

\section{Conventional Cosmological Constraints}\label{sec:reheat}

This paper will address cosmological histories where the reheat temperature, $\TRH$, is larger than the susy breaking scale, $F$.  With high reheat temperatures there are many other conventional constraints that must be avoided.  Models of susy breaking such as (low scale) Gauge Mediation \cite{Dine:1981za, Dimopoulos:1981au,Alvarez-Gaume:1981wy,Dine:1993yw,Dine:1995ag,Dine:1994vc}, Split Susy \cite{Arkani-Hamed:2004fb,Arkani-Hamed:2004yi,Giudice:2004tc,Wells:2003tf}, or AMSB \cite{Randall:1998uk} can avoid these constraints while having $\TRH>F^{1/2}$ and in these situations there is a new constraint that will be discussed in later sections. 

\begin{description}
\item[Gravitino Problem]  
In many models with low scale susy breaking the gravitino is the LSP ($m_{3/2}\sim F/M_{pl}$).  There is no bound on the reheat temperature if the gravitino mass is smaller than $1\kev$, corresponding to $F^{1/2}\sim 100-1000\tev$.  Due to their small mass, these gravitinos do not over-close the universe.  Another situation where a high reheating temperature is allowed is the case where $m_{3/2}\gtap 10\,\tev,$ in which case the gravitino is sufficiently unstable as to decay before big bang nucleosynthesis  \cite{Weinberg:1982zq}.  
\item[Polonyi Problem] 
After inflation Polonyi fields are displaced a long way from the minimum of their potential.  Their coherent oscillations can dominate the energy density of the universe.  However if they decay early enough through renormalizable interactions then they do not necessarily over-close the Universe \cite{deCarlos:1993jw, Banks:1993en}.
\item[Other Moduli problem]  
The existence of light moduli is highly model dependent.  There are known models where all the moduli are stabilized at a high scale and essentially decouple from the cosmology\cite{DeWolfe:2005uu}. 
\end{description}

Thus in principle it is possible to reheat to temperatures above $F^\half$ and not suffer from other cosmological problems.  In many situations with metastable susy breaking vacua, the barrier height is given parametrically by $F^\half$ and the issue of thermally activating transitions to the true minimum is potentially an issue.

\section{Supersymmetric QCD}
\label{Sec: ISS}

In \cite{Intriligator:2006dd} it was shown that for sufficiently small mass of the electric quarks in  $SU(N_C)$ SQCD, there is a long lived metastable vacuum.  From the low energy (magnetic) point of view, the supersymmetric vacua are created by an irrelevant operator attaining the same size as a relevant operator and exist at large field values relative to the metastable vacuum.  

At the high temperatures following reheating, the vacuum of the theory lies in neither the metastable vacuum nor the various supersymmetric vacua, but rather at the minimum of the free energy.  As the universe cools the minima of the free energy develop towards the metastable and supersymmetric vacua, making it possible to determine into which vacuum the universe evolves. 
 
The microscopic (electric) theory consists of asymptotically free $\mathcal{N} =1$ supersymmetric $SU(N_C)$ QCD with $N_F$ massive flavors.  There is an $SU(N_F)_L\times SU(N_F)_R$ approximate flavor symmetry 
with the quarks transforming as
\begin{eqnarray}
Q \sim ( \square_{N_C}, \square_{N_{F\hspace{0.02in}L}})
\hspace{0.2in}
Q^c \sim ( \overline{\square}_{N_C}, \overline{\square}_{N_{F\hspace{0.02in}R}}).
\end{eqnarray}
The  quark mass breaks the global symmetry with a superpotential
\begin{eqnarray}
W_e=m \tr QQ^c,
\end{eqnarray}
where the masses $m$ are taken to be degenerate for simplicity. 

The electric theory goes strong at a scale $\Lambda$.  Below this strong-coupling scale  the system may be described by an IR-free dual gauge theory provided $N_C < N_F <\frac{3}{2}N_C$ \cite{Seiberg:1994pq}.  The macroscopic (magnetic) theory is an $SU(N)$ gauge theory ($N = N_F - N_C$) with $N_F$ magnetic quarks $q$  and $q^c$ and has a Landau pole at $\Lambda$ and runs free in the IR. The magnetic quarks  have the same approximate $SU(N_F)_L \times SU(N_F)_R$ flavor symmetry
\begin{eqnarray}
q \sim ( \square_{N}, \overline{\square}_{N_{F\hspace{0.01in}L}})
\hspace{0.2in}
q^c \sim ( \overline{\square}_{N}, \square_{N_{F\hspace{0.01in}R}}).
\end{eqnarray}
There is also an additional gauge singlet superfield,  $M,$ that is a bi-fundamental of the flavor symmetry:
\begin{eqnarray}
M \sim ( \square_{N_{F\hspace{0.01in}L}}, \overline{\square}_{N_{F\hspace{0.01in}R}}).
\end{eqnarray}
The infrared theory is IR free with $N_F > 3N.$ The tree-level superpotential in the magnetic theory is given by
\begin{eqnarray}
W_m=y\,\tr\,q Mq^c-\mu^2 \tr\,M,
\end{eqnarray} 
where $\mu^2\sim m\Lambda.$
The $F$-terms of $M$ are 
\begin{eqnarray}
F^\dagger_{M_i^j} = y\, q^a_i q_{a}^{c\,j} - \mu^{2} \delta^j_i,
\end{eqnarray}
which cannot vanish uniformly since $\delta^j_i$ has rank $N_{F}$ but $  q^a_i q_{a}^{c\,j}$ has rank $N_F - N_C < N_F.$ As such, supersymmetry is spontaneously broken in the magnetic theory by the rank condition.
This susy-breaking vacuum lies along the quark direction
\bea 
&\vev{M}_{\text{ssb}} =0  \hspace{3mm} &\vev{q}_{\text{ssb}}=\vev{q^c}_{\text{ssb}}\sim N \mu \unit_{N} 
\eea
with $\vev{F_M} \sim \mu^2$.

When  the meson vev is turned on, the quarks decouple and the magnetic theory becomes pure $SU(N)$ super-Yang Mills.  This theory has a dynamically generated strong coupling scale, $\Lambda_m(M)$, given by
\begin{eqnarray}
\Lambda_m(M)  = M \left( \frac{M}{\Lambda}\right)^{\frac{a}{3}},
\end{eqnarray}
with $a=\frac{N_F}{N}-3,$ a strictly positive quantity when the magnetic theory is IR free.  Here, for simplicity, we have taken the meson vev to be proportional to the identity, i.e. $\vev{M}\sim M\unit_{N_F}$.  Gaugino condensation at this scale leads to an ADS superpotential \cite{Affleck:1983mk} for $M$. At temperatures above $\Lambda_m$ the contributions to the superpotential from gaugino condensation disappear.  Below the mass of the quarks this additional nonrenormalizable contribution obtains the form \cite{Affleck:1983mk} $W_{\text{det}}=\left(\frac{\det M}{\Lambda^{N_F-3N}}\right)^{\frac{1}{N}}$.  Thus the complete superpotential in the magnetic theory is given by  
\begin{eqnarray}
W=-\mu^2 \tr M + y \tr q Mq^c + \left(\det M\right)^{\frac{1}{N}} \Lambda^{-a}.
\label{eqn:superpot}
\end{eqnarray}
Interpreted physically, $a$ characterizes the irrelevance of the determinant superpotential.  For instance, in the $M \sim \eta\unit$ direction the superpotential behaves as
\begin{eqnarray}
W \sim - \mu^2 \eta + \eta^{3+a} \Lambda^{-a} .
\end{eqnarray}
The complete superpotential admits a susy-preserving solution of $F_M=0$; however because the determinant superpotential is an irrelevant operator, this vacuum is very distant from the origin (and also the metastable vacuum):
\be
\vev{M}_{\text{susy}} = \mu \left(\frac{\Lambda}{\mu}\right)^\frac{a}{2+a} \unit_{N_F}  \hspace{2mm} \vev{q}_{\text{susy}}=\vev{q^c}_{\text{susy}}=0.
\ee

Around the origin the potential goes as
\begin{eqnarray}
V \sim - \mu^2\left(y  q q^c + \eta^{2+a} \Lambda^{-a}\right).
\end{eqnarray}
Due to the irrelevance of the determinant superpotential, the meson direction rolls off slower than the magnetic quarks which are standard tachyons around the origin.

The rate per unit volume of  bubble formation of the true vacuum in the zero temperature case is
\be
\Gamma\sim \mu^{4} \exp(-S_4),
\label{eqn:zerotemprate}
\ee
where $S_4$ is the four dimensional Euclidean bounce action.
Calculating the bounce action for a general potential is only possible numerically or in the thin-wall approximation \cite{Coleman:1977py,Callan:1977pt,Coleman:1980aw}. 
However, for square or triangular potentials the solutions are known exactly \cite{Duncan:1992ai}.  Along the meson direction our potential is well approximated by a square potential barrier and the 
bounce action is approximately
\be
S_4\sim 2 \pi^{2} \frac{ \Delta \eta_{\text{susy}}^4}{(V_{\text{peak}}-V_{\text{susy}})}.
\ee
Only bubbles whose radius is greater than some critical value $R_c$ will grow and cause a transition to the true minimum.  It was shown in \cite{Intriligator:2006dd} that $S_4\sim \left(\frac{\Lambda}{\mu}\right)^{\frac{4a}{2+a}},$
which can be made arbitrarily large--and the false vacuum long lived--by taking $\mu\ll \Lambda$.
Ensuring that no transition to the supersymmetric minimum has occurred during the lifetime of the Universe (i.e., that the lifetime of the nonsupersymmetric universe exceeds 14 Gyr) places a constraint on the
theory
\begin{eqnarray}
\label{Eq: Lifetime Constraint}
\frac{a}{a+2} \log \frac{\Lambda}{\mu} \gsim 0.73 + 0.003 \log\frac{\mu}{\tev} + 0.25 \log N.
\end{eqnarray}
This is a very weak constraint (amounting to $(\Lambda/\mu)^{a/(a+2)}\gsim 2$) .  Of course, keeping the K\"{a}hler corrections under control entails
$\mu/\Lambda\ll 1,$ and in addition it is the running between $\Lambda$
and $\mu$ that makes the Yukawa and gauge couplings perturbative.

\subsection*{The Finite Temperature Potential}
\label{Sec: Finite Temp}

In the previous section the zero temperature phase diagram was illustrated.  In this section the finite temperature phase diagram is discussed using the  one loop, finite temperature effective potential corrections to (\ref{eqn:superpot}); they are  \cite{Dolan:1973qd}
\begin{eqnarray}
\delta V(\phi,T) & =  \sum_{\alpha,\text{boson}}\left( -   \frac{\pi^2}{90} T^4 +  \frac{1}{24} m_\alpha^2(\phi) T^2 + \cdots \right)   +\sum_{\alpha,\text{fermion}}\left(-  \frac{7}{8} \frac{\pi^2}{90} T^4 +  \frac{1}{48} m_\alpha^2(\phi) T^2 + \cdots\right)\nonumber
\end{eqnarray}
At sufficiently high temperatures, the minimum of the free energy of the system prefers
to have the maximum number of light degrees of freedom.
At low temperatures, these modes decouple and the thermal contributions to the effective potential vanish; here the dynamics are determined by the zero temperature potential.

If the system is in thermal equilibrium at a temperature $\Lambda$, the highest temperature where the magnetic theory is valid, then the minimum of the thermal potential lies at the origin, i.e. $\vev{q}=\vev{q^c}=\vev{M}=0$.  As the Universe cools other minima of the potential will develop elsewhere and the origin will eventually become unstable, leading to a phase transition to either the supersymmetric vacua or the metastable vacuum. The phase transitions to the metastable and supersymmetric minima are explored in the next two sections.

\subsection*{The Transition to the Metastable Vacuum}

As we noted before, the magnetic quarks are tachyons around the origin with a negative mass squared of $-\mu^2$.  
As the quark field acquires a vacuum expectation value, the meson and anti-quark marry and acquire a mass, causing the one loop correction to the thermal effective potential to stabilize the quark (and similarly for the anti-quark) at the origin of field space at sufficiently high temperatures. 
We parametrize the fields in the quark direction as
\be
q=q^c= \frac{1}{\sqrt{2N}}\left( \begin{array}{cc} \xi \unit_N& 0\end{array}\right),
\ee
where the quark vevs are degenerate since this is a $D$-flat direction and preserves a $SU(N)_{D} \times SU(N_F - N)$ flavor symmetry.  The non-$D$-flat directions 
cause fields to acquire more mass and are stabilized more quickly than the $D$-flat direction.  This makes the $D$-flat directions suitable for study.  Similar arguments apply to directions the preserve 
a smaller flavor subgroup.
The leading finite-temperature corrections to the potential are of the form $\xi^2 T^2;$ explicitly we find
\begin{eqnarray}
V= N  \left(\frac{y \xi^2}{N^2}-\mu^2\right)^2 -  c_0N_F^2 T^4 
 + (c^{(g)}_1 g^2 + c^{(y)}_1 y^2) N \xi^2 T^2+ \ldots,
\label{eqn:finitetquarkpot}
\end{eqnarray}
where $c^{(g,y)}_1\sim \mathcal{O}(1)$ and $c_0$ is a pure number.  This is essentially identical to the standard renormalizable Higgs potential at finite temperature that develops a second-order (or perhaps weakly first order) phase transition at critical temperature 
\begin{eqnarray}
T_c^{\text{ssb}}\sim \frac{\mu}{y^{\half} N},
\end{eqnarray}
in which the origin becomes unstable and the minimum moves approximately smoothly to $\vev{q}\ne 0$.
There is no impediment to making this transition so long as the quarks and mesons are not at the susy preserving vacuum at $T_c^{\text{ssb}}$.

\subsection*{The Transition to the susy Vacuum} 

The phase transition to the supersymmetric vacuum is more interesting.   Any transition to a lower minimum away from the origin must be first order and occur through bubble nucleation.  The meson vev in this direction breaks $SU(N_F) \times SU(N_F) \rightarrow SU(N_F)_{D}.$ We parametrize the fields in the meson direction via
\begin{eqnarray}
M=\frac{\eta\unit}{\sqrt{N_F}}
\end{eqnarray}
where other directions cause fields to acquire larger masses and therefore are stabilized more
readily.  We find the finite temperature potential in this direction to be given by
\be
 V=\begin{cases}
\mu^4  + c_1\,y^2  N \eta^2 T^2-c_0(N N_F +N^2) T^4 + \cdots & \text{\hspace{-3mm} $T\ge 
\Lambda_m(\eta)$}\\
N \Lambda^4 \left|\left(\frac{\eta}{\sqrt{N_F}\Lambda}\right)^{2+a}-\frac{\mu^2}{\Lambda^2}\right|^2  \\
\hspace{0.10in} + c_1\,y^2  N \eta^2 T^2-c_0N N_F T^4 + \cdots &   \text{\hspace{-3mm} $T\ge y \eta$} \\
N \Lambda^4 \left|\left(\frac{\eta}{\sqrt{N_F}\Lambda}\right)^{2+a}-\frac{\mu^2}{\Lambda^2}\right|^2 & \text{\hspace{-3mm} $T < y\eta$}
\end{cases}
\label{eqn:finitetmesonpot}
\ee
where $c_0$ and $c_1$ are dimensionless pure numbers.

Since $a>0$ the zero-temperature potential, near the origin, behaves as
\begin{eqnarray}
V \simeq - \mu^2 \Lambda^{-a} \eta^{2+a}.
\end{eqnarray}
The additional contribution from the quark superfields can stabilise the origin even at very low temperatures.  This means any transition to a lower minimum in the meson direction will take place via tunneling.  The existence of fields that become light at the origin of the Polonyi field is crucial to the stabilization.  Models with several dimensionful scales, such as the \oraf model of Sec. \ref{Sec: ORaif}, lack this feature and do not have a stable origin at finite temperature.

\begin{figure}
\centerline{\includegraphics{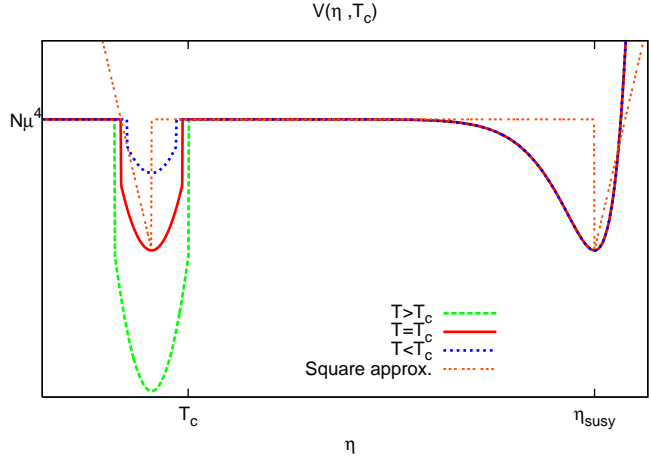}}
\caption{Cartoon of the potential in the meson direction, for temperatures at, above and below the critical temperature, $T_c^{susy}$, here $\eta_{\text{susy}}=\left(\mu^2\Lambda^a\right)^{\frac{1}{2+a}}$.  Also plotted is the square approximation to the potential used for calculating the bounce action.}
\label{fig:mesonpot}
\end{figure}

The critical temperature, $T_c^{\text{susy}}$, is the temperature at which the second minimum  of the potential is degenerate with the vacuum at the origin; see Figure \ref{fig:mesonpot} for a plot of the potential in the meson direction.  Since the second minimum is determined by irrelevant interactions, it is always at large field values in comparison with $\mu$.    This means that around this second minimum the quarks are particularly heavy and they no longer contribute to the potential, whereas near the origin they are the dominant contribution to the potential.  The degeneracy of the two minima occurs at 
\be
T_c^{\text{susy}}\sim \left(N_F+N\right)^{-\frac{1}{4}}\mu.
\ee
Although our potential (\ref{eqn:finitetmesonpot}) has a discontinuity at $y \eta=T_c^{\text{susy}},$ in reality there would be a smooth transition between the two limits.  At this value of $\eta$ the peak of the potential barrier is of height $V_{\text{peak}}\sim N \mu^4$.

The bubble nucleation rate of the susy vacuum at finite temperature is
\be
\Gamma \sim T^4 e^{-S_3/T}
\label{eqn:tunnelprob}
\ee
where $S_3$ is the three dimensional bounce action.  Using the square barrier again and extending the result to three dimensions, we find that the bubble action is given by
\be
S_3 \sim \frac{4 \pi}{3 \sqrt{2}} \frac{\Delta\eta_{\text{susy}}^3}{\left[(V_{\text{peak}}-V_{\text{susy}})^{\frac{1}{4}} - (V_{\text{peak}} - V_{0})^{\frac{1}{4}}\right]^{2}}.
\ee
where $V_{0}$ is the potential at the origin of field space. The transition to the true minimum is dominated by bubbles of a size $R_c$
\begin{eqnarray}
R_c \sim  \frac{\Delta \eta_{\text{susy}} \Delta V^\half}{\delta V}
\end{eqnarray}
where $\Delta V\simeq \mu^4$ is the height of the barrier and $\delta V$ is the potential difference between the true minimum and the false minimum, given by
\begin{eqnarray}
\delta V \simeq \mu^4 \left(1 -  \left(\frac{T}{T_c^{\text{susy}}}\right)^4\right).
\end{eqnarray}
At temperatures beneath $T_c^{\text{ssb}}$, the quarks roll off to the
metastable minimum, but this does not make a parametric difference
in the arguments above.  The transition to the supersymmetric vacuum is dominated by bubbles of size $R_c$ as
bubbles smaller than $R_c$ shrink and larger are much larger action.    At temperatures below $T_c^{\text{susy}}$ and above $R_c^{-1}$ the tunneling rate is the finite temperature result (\ref{eqn:tunnelprob}).  At temperatures below $R_c^{-1},$  bubbles of radius $R_c$ may no longer be thermally nucleated and the tunneling rate changes into the zero temperature result (\ref{eqn:zerotemprate}).  The radius of the critical bubble, $R_c$, is considerably smaller than the Hubble length and so all gravitational effects can be ignored.  The transition temperature is given by
\begin{eqnarray}
T_{\text{3d}\rightarrow\text{4d}} \simeq  \mu \left(\frac{\Lambda}{\mu}\right)^{-\frac{a}{2+a}},
\end{eqnarray}
though  the rate shuts off well before this low temperature.   
However, since $\delta V$ at the transition temperature increases the the action for these bubbles, the largest contribution to bubble nucleation comes from temperatures at $T = T_c^{\text{susy}}/3,$ where
\be
\log \frac{\Gamma}{\mu^4} \simeq - \frac{9 \pi}{\sqrt{2N}} \left(\frac{\Lambda}{\mu}\right)^{\frac{3a}{2+a}}.
\ee
The crucial feature of this result is that the quantity that guarantees the longevity of the Universe also protects the cosmological evolution:
\begin{eqnarray}
\left(\frac{\Lambda}{\mu}\right)^{\frac{a}{2+a}}  \gg 1.
\end{eqnarray}
Thus the system evolves towards the metastable vacuum so long as the Universe is cosmologically long lived.

In order to avoid tunneling into the true vacuum during the evolution of the universe, we require the fraction of space remaining in the false vacuum \cite{Guth:1981uk} at the end of thermal tunneling to be of order unity, corresponding to the constraint
\be
\Gamma(T) \,a^{3}(T)\, \mathcal{V} \Delta t \simeq 0 ,
\ee
where $\mathcal{V}$ is the  spatial size of our universe today; $a^3(T)$ is the scale factor blueshifting this volume to the appropriate size at the time the thermal transitions are active; and $\Delta t \sim H^{-1}(T) \sim M_P/T^2$ is the duration of this period.
This constraint is satisfied for 
\begin{eqnarray}
\label{Eq: Cosmology Limit}
\frac{a}{a+2}\log \frac{\Lambda}{\mu} \gtrsim  0.64 - 0.010 \log\frac{\mu}{\tev}  + 0.17 \log N . 
\end{eqnarray}
This  constraint is somewhat weaker than the condition that the lifetime of the metastable vacuum is greater than 14 Gyr found in (\ref{Eq: Lifetime Constraint}).  Therefore if the metastable vacuum is sufficiently long lived, Big Bang cosmology will evolve towards the metastable vacuum rather than the true minimum.

\subsection*{Recap}

To summarize the calculations of the previous section, at temperatures above $\OO(\mu)$ the minimum of the free energy is at the origin of field space and the full $SU(N)$ gauge and $SU(N_{F\hspace{0.01in}L})\times SU(N_{F\hspace{0.01in}R})$ global symmetry is restored.  Beneath this temperature both the supersymmetric and susy breaking vacua become minima of the free energy; however, the transition to the susy breaking vacuum is a second order phase transition that occurs rapidly, while the transition to the susy preserving vacua is first order.  
At the origin of field space there are many light states that acquire a large mass near the susy preserving minimum.  Creating a bubble of the true vacuum requires decoupling these light fields from the theory near the origin.  This fluctuation is entropically costly and makes the transition strongly first order.  This is to be contrasted with the transition to the metastable minimum where no fields acquire masses parametrically larger than the temperature.
The result is that the Universe quickly evolves into the susy breaking vacuum, with the supersymmetric vacua only reached through thermally activated bubble nucleation.   The thermal bubble nucleation rate is closely related to the zero temperature rate, so that the first-order phase transition may be adequately suppressed by ensuring that the lifetime of the Universe is sufficiently long.  Which of the two constraints is stronger turns out to be related to details of the model, but they will always be closely related for the following reason:
The present age of the Universe, the weak scale, and the cosmological constant are related via the Planck scale and constitute the largest contribution to the constraints. The actions for the zero- and finite-temperature bubbles are also closely related, so the differences in the log of the rates lie in $\OO(1)$ numbers.  
\subsection{Related Models}

There exist a variety of deformations of the ISS model that exhibit similar cosmological evolution. 
One example is to add to the electric theory an adjoint chiral superfield, $\Phi$ with the superpotential
couplings
\begin{eqnarray}
W_{el} = \mu \tr Q Q^c+ \frac{\kappa}{3} \tr \Phi^{3} + \frac{m_{\Phi}}{2} \tr \Phi^{2} + \lambda \tr Q \Phi Q^{c}
\end{eqnarray}
Supersymmetric QCD with these additions constitutes the $k = 2$ Kutasov model \cite{Kutasov:1995ve,Kutasov:1995np,Kutasov:1995ss}; it is asymptotically free with stable vacua for $N_{C}/2 < N_{F} < 2 N_{C}.$ 

This microscopic theory has a dual magnetic description in terms of a $SU(N)$ gauge theory with $N= 2 N_{F} - N_{C}$,  $N_{F}$ magnetic quarks $q, q^{c},$ a magnetic adjoint $Y,$ and two gauge singlet mesons $M_{1}, M_{2},$ where $M_{1} = QQ^{c},$ $M_{2} = Q \Phi Q^{c}.$ The dual magnetic superpotential of the theory is
\begin{eqnarray}
\nonumber
W_{mag} &=& \frac{g_{Y}}{3} \tr Y^{3} + \frac{m_{Y}}{2} \tr Y^{2} -  m_{1}^{2} \tr M_{1} - m_{2}^{2} \tr M_{2} \\
 && + \tr (h_{1} M_{1} q q^{c} + h_{2} M_{2} q q^{c} + h_{3} M_{1} q Y q^{c}) 
\end{eqnarray}

The corresponding condition on $N$ for an IR free magnetic theory is $0 < 2N < N_{F}.$ As in the ISS theory, supersymmetry in this model will be broken by the rank condition in the $M_{2}$ F-term: $F_{(M_{2})^{i}_{j}} = h_{2} q^{i} q^{c}_{j} - m_{2}^{2} \delta^{i}_{j}.$ Since $N_{F} > N,$ not all of these F-terms may be set to zero, and supersymmetry is broken, with $M_2$ acquiring an F-term vev and the dual squarks taking on vacuum expectation values.

 When the $M_{2}$ meson obtains a vev, the magnetic quarks decouple and the adjoint may be integrated out. As in the ISS model, the theory has a dynamically generated strong coupling scale $\Lambda_{m}(M_{2}),$ given by
\begin{align*}
\Lambda_{m}(M_{2}) = m_{Y}^{1/3} M_{2}^{2/3} \left( \frac{M_{2}}{\Lambda} \right)^{\frac{a}{3}},
\end{align*}
where now $a = \frac{N_{F}}{N} - 2,$ a strictly positive quantity when the magnetic theory is IR free. The supersymmetry-preserving vacuum is created again by gaugino condensation at this strong-coupling scale, and is given by an ADS superpotential of the form $W_{\text{det}} = m_{Y} \left(  \frac{\det M_{2}}{\Lambda^{N_{F} - 2N}} \right)^{\frac{1}{N}}.$ The supersymmetric vacuum is at a large vev in the $M_{2}$ meson direction relative to the susy breaking vacuum because the ADS superpotential looks like an irrelevant operator from the low energy point of view.

The metastable vacuum is  long-lived because leaving it requires tunneling to a distant point in field space \cite{Amariti:2006vk}. The zero-temperature tunneling parameter in this theory for transitions into the supersymmetric vacuum is given by
\begin{eqnarray}
S_{4} = 2 \pi^{2} \left( \frac{m_{2}}{m_{Y}} \right)^{\frac{4}{1+a}} \left( \frac{\Lambda}{m_{2}} \right)^{\frac{4a}{1+a}}.
\end{eqnarray}
The longevity of the metastable vacuum is allowed by $\frac{\Lambda}{m_{2}} \gg 1,$ provided $m_{2} \sim m_{Y}.$   For a realistic model to have a  lifetime greater than 14 Gyr requires
\begin{eqnarray}
\frac{a}{a+1} \log \frac{\Lambda}{m_{2}} \gsim 0.73 +0.003 \log \frac{m_{2}}{\TeV} + \frac{1}{a+1} \log \frac{m_{Y}}{m_{2}} + 0.25 \log N,
\end{eqnarray}
which mimics ISS with an additional dependence on the mass ratio $m_{Y}/m_{2}.$

The phase transition transition into the susy-breaking minimum is second-order, while the transition into the supersymmetric minimum is first-order if $a >1$ and  second-order (or weakly first-order) if $0 < a \leq 1$. 
The finite-temperature tunneling parameter for tunneling into the susy vacuum in the square barrier approximation is
\begin{eqnarray}
\frac{S_{3}}{T} = \frac{4 \pi}{3 \sqrt{2N}} \frac{m_{2}}{T} \left( \frac{m_{2}}{m_{Y}} \right)^{\frac{3}{1+a}} \left( \frac{\Lambda}{m_{2}} \right)^{\frac{3a}{1+a}} \frac{1}{[1-T/T_{c}]^{2}}.
\end{eqnarray}
leading to a finite-temperature bound on the theory of
\begin{eqnarray}
\frac{a}{a+1} \log \frac{\Lambda}{m_{2}} \gsim 0.64 - 0.010 \log \frac{m_{2}}{\TeV}  + 0.17 \log N + \frac{1}{a+1} \log \frac{m_{Y}}{m_{2}}.
\end{eqnarray}

Thus the generic constraints Kutasov models follow the same lines as the ISS model--the longevity of the universe is sufficient to guarantee the selection of the metastable vacuum. Perhaps more interestingly, this theory admits a class of solutions ($0 < a \leq 1,$ or $2N < N_{F} \leq 3 N$) for which the determinant superpotential is marginal or relevant, and thus the phase transition into the susy vacuum may be second-order. Nonetheless, the supersymmetric vacuum exists far out in field space; the presence of light fields around the origin ensures that the theory will be stabilized against transitions into the supersymmetric vacuum, provided the cosmological longevity bound is satisfied by the parameters of the theory.

\subsection{Gravitationally Stabilized Gauge Mediation}
 
A less promising situation is where the susy breaking minima are created
at distant values in field space where there are fewer light species than at the  origin.
 A recent example was given in \cite{Kitano:2006wz}.   In this model, a susy breaking minimum was created through irrelevant interactions of the supergravity potential.   The theory consists of a singlet chiral superfield $S$ and vector-like messenger superfields $q, q^c$ and a superpotential 
\begin{eqnarray}
W = \mu^{2} S - \lambda S q q^{c} + m^3
\end{eqnarray} 
and a K\"{a}hler potential 
\begin{eqnarray}
K = S^{\dagger} S - \frac{(S^{\dagger}S)^{2}}{2\sqrt{3} \tilde \Lambda^{2}}  + q^{\dagger} q + q^{c\,\dagger} q^c,
\end{eqnarray} 
where the constant $m^3$ is relevant to the cancellation of the cosmological constant and $\tilde \Lambda$ is the cut-off of the low energy theory.

There exists a susy-preserving minimum lying in the $q, q^c$ direction at
\begin{eqnarray}
\nonumber
\vev{q}_{\text{susy}}=\vev{q^c}_{\text{susy}} \simeq \frac{\mu}{\sqrt{\lambda}} + \mathcal{O} \left( \frac{m^{3}}{\lambda M_{P}^{2}} \right) \quad 
\vev{S}_{\text{susy}} \simeq \mathcal{O} \left( \frac{m^{3}}{\lambda M_{P}^{2}} \right)  ,
\end{eqnarray}
where the shifts of $\mathcal{O} \left( \frac{m^{3}}{\lambda M_{P}^{2}} \right)$ due to gravitational interactions are negligible. The theory in this direction resembles the quark direction of the ISS model, albeit with a supersymmetry-preserving minimum.  This potential develops a second-order (or weakly first-order) phase transition into the supersymmetric minimum at a critical temperature $T_{c}^{\text{susy}} \sim \frac{\mu}{\sqrt{\lambda}}.$

In the $S$ direction, a susy-breaking minimum is created away from the origin through the  $M_P$ suppressed operators with $\mu$ and $m$ balancing each other to cancel the cosmological constant
$m^{3} \sim  \mu^{2} M_{P}$.   $S$ acquires an F-term vev, $F_S\simeq \mu^2,$ and the potential in this direction is of the form
\begin{eqnarray}
V_{0}(S) \approx \frac{2}{\sqrt{3}} \mu^{4} \left| \frac{S}{\tilde\Lambda} - \frac{\tilde \Lambda}{ M_{P}} \right|^{2} + ...
\end{eqnarray}
The minimum lies far from the origin relative to the susy preserving one at
\begin{eqnarray}
\vev{S}_{\text{ssb}} \sim \frac{\tilde{\Lambda}^2}{ M_p} \hspace{0.2in} \vev{q}_{\text{ssb}}=\vev{q^c}_{\text{ssb}}=0.
\end{eqnarray}
The bubble action for tunneling from the susy breaking vacuum to the susy preserving one is given parametrically
by
\begin{eqnarray}
S_4 \sim \left( \frac{\tilde\Lambda^2}{\mu M_p}\right)^4 .
\end{eqnarray}
The bubble action is large if $\tilde \Lambda \gsim 10^{8}\,\tev\times \left(\mu/\tev\right)^\half$ . 

If this sector is thermalized at a temperature greater than $F_S^\half$, finite-temperature corrections tend to stabilize the theory at the origin at high temperatures, away from the susy-breaking vacuum. 
The very same features that served to guarantee the thermal evolution of SQCD-based theories into their susy-breaking vacua prevents the evolution of this gravitationally-stabilized theory into its supersymmetry breaking vacuum.  
At a temperature $T_{c} \sim \frac{\mu}{\sqrt{\lambda}}$ there is a second order phase transition to the susy preserving vacuum, while the transition to the susy breaking vacuum is never energetically favorable.
If this hidden sector reaches thermal equilibrium above a temperature $\OO(\mu)$, thermal evolution will take it into the supersymmetric vacuum.  

This argument obviously does not exclude the possibility of a different ansatz for the early Universe that would allow a dynamical explanation for occupation of the false vacuum, as we will discuss in Sec. \ref{Sec: Early Universe}. Nonetheless, it illustrates the utility of finite-temperature arguments in evaluating susy-breaking models, as parametric longevity of the metastable vacua at zero temperature may not be sufficient to guarantee the cosmological validity of the theory.
  
  
 \section{The Classic O'Raifeartaigh Model}
 \label{Sec: ORaif}
 
 In the previous section it was shown that reheated ISS-like O'Raifeartaigh Models always cool
 to the vacuum closest to the origin.  Furthermore, thermally activated bubbles that stimulate vacuum decay from the false minimum to the true minimum do not place a parametrically stronger constraint on the theories  than guaranteeing that the Universe lives for a sufficiently long period of time.  This came about because there were massless fields around the origin that created an entropic cost to making large excursions in field space at high temperatures.  This is not always the case and minimal O'Raifeartaigh models,  \cite{O'Raifeartaigh:1975pr}, provide counter examples that are illustrated here.
In this case the origin is not as non-perturbatively stable as ISS-like models at finite temperature.

The classic O'Raifeartaigh superpotential involves three superfields $\psi, \psi^{c}, Z$ with superpotential
\begin{eqnarray}
W = m \psi \psi^c + \lambda Z( \psi^2 - \mu^2)
\end{eqnarray}
and canonical K\"{a}hler terms.  

In this canonical O'Raifeartaigh model, the $F$-flatness conditions can not be simultaneously
satisfied and if $\mu < m$ the vacuum is given by
\begin{eqnarray}
\vev{\psi}_\ssb = \vev{\psi^c}_\ssb =\vev{Z}_\ssb=0  
\end{eqnarray}
with $F_Z=\lambda \mu^2$.  Conversely, if $\mu >m$, $\psi$ acquires a vacuum expectation value and the theory returns
to the same form as above after re-diagonalizing the fields and exchanging the role of
$\psi^c$ and $Z$.  However, apart from a relabeling,
the dynamics are identical to the first case and henceforth we will restrict ourselves to 
considering the case $\mu< m$.

To analyze this theory in the limit $m\gg\mu$ we may 
integrate out the O'Raifeartaigh fields, $\psi \psi^c$, and consider the low energy effective action of only the Polonyi field, $Z$, beneath the scale $m.$  The superpotential here is simply that of the Poloyni model,
\begin{eqnarray}
W = -\mu^2 Z.
\end{eqnarray}
Upon integrating out the O'Raifeartons, the K\"{a}hler potential picks up a one loop renormalization, given by
\begin{eqnarray}
K = \left(1 - \frac{c \lambda^2}{16\pi^2} \log( 1 + \frac{\lambda^2 |Z|^2}{m^2})\right) |Z|^2.
\end{eqnarray}
This renormalized K\"{a}hler potential stabilizes $\vev{Z}$ at the origin and is valid for large values of $Z$ until the logs have to be resummed.  At large and small field values, the scalar potential reduces respectively to the limits
\begin{eqnarray}
\nonumber
V(Z) &\simeq& \mu^4 + \frac{c\lambda^4\mu^4}{8\pi^2m^2} |Z|^2 \hspace{0.5in} Z\ll m \\
V(Z) &\simeq& \frac{\mu^4}{1 - \frac{c\lambda^2}{16\pi^2} \log \frac{ \lambda^2 |Z|^2}{m^2}}
\hspace{0.5in} Z\gg m.
\end{eqnarray}

This theory possesses a single minimum in which supersymmetry in broken. However, it is possible that whatever generates the scale $\mu\ll M_P$ also generates corrections the superpotential.
Additional contributions to the superpotential of the form
\begin{eqnarray}
\delta W =\half \epsilon \mu Z^2 
\end{eqnarray}
create a supersymmetric vacuum at field values
\begin{eqnarray}
\vev{Z}_\susy = \epsilon^{-1}\mu\hspace{0.5in} \vev{\psi}_\susy=0.
\end{eqnarray}
In this case the theory possesses two vacua, and the susy-breaking vacuum at the origin becomes metastable. 
Adding the mass term for the Polonyi field results in a potential
\begin{eqnarray}
V(Z) = \mu^4 \frac{|1 - \epsilon Z/\mu|^2}{\partial_{Z}\partial_{\bar{Z}} K }.
\end{eqnarray}
This retains a local minimum near the origin due to the K\"{a}hler corrections when $\epsilon \ll \lambda^2/16\pi^2.$ The potential peaks around the value, for simplicity from now on we set $c=1$,
\begin{eqnarray}
Z_{\text{peak}} = \frac{\lambda^2}{16\pi^2}\frac{\mu}{\epsilon}.
\end{eqnarray}
Notice that the peak is determined at large field values relative to
$m,$ where the only dimensionful variation of the scale is on the
order $\mu/\epsilon$.
The height of the potential at its peak is roughly
\begin{eqnarray}
\Delta V \approx \frac{\lambda^2}{8\pi^2} \mu^4 .
\end{eqnarray}
In this case, the lifetime of the metastable vacuum of the origin 
may be estimated using a triangular potential.  With the parameters
of this potential, the action is dominated by a bubble in which the field tunnels to the top of the potential and then rolls down to the supersymmetric minimum.
The tunneling action for this 
non-thin wall bubble is approximated
by 
\begin{eqnarray}
S_4\simeq 2\pi^2 \frac{ (Z_{\text{peak}})^4}{\Delta V} = 2\pi^2 \left( \frac{ \lambda^2}{16 \pi^2}\right)^3 \epsilon^{-4}.
\end{eqnarray}
By making $\epsilon$ small enough, the lifetime of the universe can be made arbitrarily long.

Above a temperature $T \sim m$, the Yukawa interactions stabilize all three fields at the origin of field space.  For temperatures beneath $m$, $\psi$ and $\psi^c$ decouple.  
Whether the O'Raifeartaigh fields are stable and how efficiently they annihilate is a model-dependent question.  In the minimal model under consideration, they are stable and can only annihilate into $Z$s through the coupling $\lambda$. 
For simplicity consider adding fields to the theory that cause $\psi$ to decay quickly. \footnote{If instead the O'Raifeartons are stable there can be a significant number density around contributing to the free energy; we thank Takemichi Okui for pointing this out.  This can lead to a stabilizing effect similar to the dynamics of MaVaNs\cite{Fardon:2003eh}. 
The stabilization by massive particles is not as effective as having light states.} 

At low energies, $Z$ only has irrelevant interactions and thus, aside from the $-T^4$ contribution to the thermal effective potential (which does not depend on $Z$),  the one-loop free energy is simply the zero-temperature potential.  There are no additional interactions that hold $Z$ at the susy-breaking minimum.  Since the barrier height is set by $\mu^4$ and can be much less than $m^4$, it is possible for the fields to undergo large thermally-induced excursions in field space to the top of the barrier and then slide down unsuppressed.     

The finite temperature bubble nucleation rate is dominated by the
three-dimensional bubbles as long as
\begin{eqnarray}
T \gsim 4\pi \lambda \epsilon \mu
\end{eqnarray}
This bubble nucleation rate is given by
\begin{eqnarray}
\label{Eq: FT OR Bubble}
S_3/T = 4\pi \frac{ (\delta Z)^3}{ (\delta V)^\half T} = 4\pi \left(\frac{\lambda}{4\pi}\right)^5\epsilon^{-3} \frac{\mu}{m}
\end{eqnarray}
where the temperature $T = m$ has been used as the first temperature at which the transition is kinematically allowed, and where the dominant contribution to the tunneling rate occurs.  Whereas the condition on the longevity of the Universe only placed a limit on $\epsilon$, the finite temperature rate places an additional limit on $\mu/m$ that arises because it is possible to reheat parametrically above the barrier height.  

The parametrically different constraint coming from thermal stability is more stringent than that coming from the zero temperature lifetime.  This is a feature of not having massless particles at the origin until temperatures of order $F^\half$.  The absence of this entropic contribution to the free energy allows the thermal stimulation of bubbles that require large excursions in field space.   


\section{Moduli Trapping and Thermalization}
\label{Sec: Early Universe}

Thus far it has been assumed that the hidden sector has thermalized.  
In this section the thermalization of ISS-like models is briefly discussed.  
It is difficult to make exact statements about thermalization since they depend on
the details of how the susy breaking is transmitted to the MSSM; however, some details
are more universal and are explored here. 
A plausible scenario to start out of thermal equilibrium is if there is a high scale of inflation where the inflaton decays only to the MSSM and the reheat temperature is significantly lower than the Hubble constant during inflation.  
In this case the initial conditions for the FRW phase of the Universe are $\vev{Q} \sim H\gg T,\Lambda$ which can easily result in the fluctuations of $Q$ being the dominant form of energy in the Universe.    The oscillations of $Q$ will not last a long time because as the quarks oscillate, they will go through the origin of field space where the vectors bosons become light and particle production will ensue  \cite{Kofman:1997yn,Kofman:2004yc,Dine:1983ys}.
With the exception of the small electric quark mass, the electric theory has a moduli space in the $D$-flat directions; squarks with large vevs will initially oscillate along the $D$-flat directions, dumping energy into the production of vectors as they pass through an enhanced symmetry point at the origin.   The figure of merit is the non-adiabaticity parameter which is  
\begin{eqnarray}
I =   \frac{ (\Delta Q)^2}{ \langle\dot{Q}\rangle}
\end{eqnarray}
where $\Delta Q$ is the amount by which the oscillations miss the true origin of field space and $\langle \dot{Q}\rangle$ is the average value of velocity of the field as it oscillates through the origin.   In order to obtain significant particle production $I \ll 1$.   At zero temperature, as $Q$ passes close to the origin, the dynamics become best described by the magnetic theory.  Here the largest one would expect $\Delta Q$ to be is $ \sim \mu$ and  $\langle \dot{Q}\rangle\simeq m Q_0$ with $Q_0$ being the initial field amplitude.  The non-adiabaticity parameter is thus given by
\begin{eqnarray}
I = \frac{\Lambda}{Q_0}
\end{eqnarray}
where $\mu^2= m \Lambda$ has been used.
As the quarks oscillate past the origin, the vectors become massive and promptly decay into massless quarks and squarks (which are massless because they are Goldstone bosons of the broken flavor symmetry).  This results in the energy of the squark oscillations rapidly converting into radiation.  The oscillations of the squark fields are exponentially damped with a decay constant
\begin{eqnarray}
t_{\rm{damp}} \sim \frac{2 \pi}{m} \left( \frac{2 \pi}{g} \right)^{3/2}.
\end{eqnarray}
The electric quarks are quickly localized at the origin; their initial energy goes into radiation which reheats the Universe to a high temperature.  Thermal equilibrium is reached at a temperature $T \gg \mu$.  

Similar arguments go through for the case where the initial field oscillations are large compared to $\mu$ but small enough that the magnetic theory is always the valid description.  Analytically this is more difficult to show, but numerically, meson oscillations pass near the origin after several oscillations so long as the mesons have enough kinetic energy to reach the origin.  Both the quark and meson oscillations quickly damp and the system reaches thermal equilibrium with the fields localized at the origin.  If the hidden sector has relevant couplings to the MSSM (e.g. in gauge mediated models), then produced particles will thermalize with the visible sector quickly.

The important feature of these models is that there are massless fields around the origin which greatly enhance the effects that dynamically trap fields around the origin.   The same effects that lead to trapping and thermalization also make the origin the minimum of the free energy at high temperatures; this subsequently evolves into the nearest vacuum as the temperature cools.    These arguments apply to the Kutasov models as well.
In models like the retrofitted, ``classic'' O'Raifeartaigh model discussed in Sec. \ref{Sec: ORaif} where the O'Raifeartaigh fields are massive around the origin, trapping effects are less active.  This means that finding desirable initial conditions requires more assumptions. 

\subsection{Moduli Trapping in Gravitationally Stabilized Gauge Mediation}

The situation for gravitationally stabilized gauge mediation is more complicated.    This is because if $S$ is displaced far from the origin, it oscillates about the susy breaking minimum and not the origin of field space where the messengers, $q$, are light.  This means that the non-adiabaticity parameter for the quarks in the oscillating $S$ field  with an amplitude, $S_0$, is given by
\begin{eqnarray}
I= \frac{\vev{S}_\ssb^2\sin^2\theta}{ S_0 m_S} .
\end{eqnarray}
where $\theta$ is the angle of the inital $S$ vev relative to the location of the susy preserving origin.  The numerator has the simple interpretation of being the impact parameter of the oscillation\footnote{We thank R. Kitano for explaining this to us.}.
In order to not be trapped at the origin $I$ should be much greater than unity.  The maximum initial displacement for $S$ is $M_p$ and relating $m_S$ and $\vev{S}_\ssb$ to the fundamental parameters
this corresponds to
\begin{eqnarray}
I =\frac{ M_P^3\mu^2}{\Lambda^5 \sin^2\theta} \gg 1
\end{eqnarray}  
The constraint on the lifetime of the Universe is stronger unless
$\sin^2\theta \ll  \Lambda/M_P$.        Thus, so long as the inflaton does not directly reheat this sector, there appears to be a viable cosmological story as to how the susy breaking minimum is reached.    The one caveat is that it has been implicitly assumed that oscillations greater than $\Lambda$ are not qualitatively different.  Since this is the scale where K\"{a}hler corrections become important, this assumption is not completely justified in the low energy effective theory.     If the oscillations are restricted to be smaller than $\Lambda$, the non-adiabaticity is even smaller and the theory is even more safe.

\section{Discussion}
\label{Sec: Conclusion}

In theories with many vacua, understanding where the Universe starts is an important question.  The string landscape typically has string-scale potential barriers and the population of these vacua usually requires mechanisms like eternal inflation.   Inside a single valley, there may be many metastable minima that are separated by sub-stringy barriers and their population may be understood through field-theory dynamics.  In this note we have addressed the simplest possible mechanism for populating vacua, through thermal evolution of the free energy.   In O'Raifeartaigh models with massless fields around the origin (e.g. ISS-like models) and with reheating to temperatures above the barrier, the thermal evolution takes the theory to the minimum closest to the origin.   If this turns out to be the metastable, susy breaking minimum, there are no additional constraints on the parameters of the theory after guaranteeing that the lifetime of this minimum is cosmologically viable.

In the case where the metastable minimum is the distant vacuum from the origin when compared to the supersymmetric vacuum, it becomes dangerous to reheat the susy breaking sector to temperatures above the barrier height. However, if at the end of inflation the fields start their motion at large vev then they may instead evolve to the far minimum.  The probability of this outcome is a detailed question depending on how the field approaches the origin since here there can be production of light states resulting in a large damping force. If there are no massless fields around the origin, like in the original O'Raifeartaigh model, there is no entropic cost to moving away.  With a high reheat temperature thermal tunneling becomes less suppressed.  The thermal evolution will lead to additional constraints on the parameters of the theory and/or the reheat temperature even after ensuring the lifetime of the Universe is cosmologically viable. 

While this work was being completed a paper taking different cosmological initial conditions was published \cite{Abeletal}.

\section*{Acknowledgements}

We thank Willy Fischler and Vadim Kaplunovsky for useful conversations near the beginning of this work.
We thank Josh Erlich, David B. Kaplan, Ryuichiro Kitano, Igor Klebanov, Markus Luty, John McGreevy, Takemichi Okui, Michael Peskin, Yael Shadmi, Steve Shenker, Yuri Shirman, Eva Silverstein, and Carlos Wagner for useful conversations.
PJF and JGW thank the University of Texas at Austin for their hospitality during the work.
NJC and JGW thank Johns Hopkins University for their hospitality during the later stages of this work.  
PJF thanks  the Aspen Centre for Physics where part of this work was completed.
JGW thanks the Galileo Institute where part of the work was performed.
PJF was supported in part by the Director, Office of Science, Office of High Energy and Nuclear Physics, Division of High Energy Physics, of the US Department of Energy under contract DE-AC02-05CH11231.
The work of NJC and JGW is supported by National Science Foundation grant PHY-9870115 and the Stanford Institute for Theoretical Physics. 
NJC was supported by an NDSEG fellowship.

\bibliographystyle{JHEP}
\bibliography{seibergcosmorefs}

\end{document}